\documentclass[12pt]{article}
\usepackage{wrapfig}
\usepackage [english]{babel}
\usepackage[applemac]{inputenc}
\usepackage[intlimits]{amsmath}
\usepackage{graphicx}
\usepackage[small]{caption}

\usepackage{setspace}
\allowdisplaybreaks
\DeclareGraphicsExtensions{.pdf,.png,.jpg,}
\usepackage{color}
\usepackage{float} 
\usepackage{slashed}
\usepackage{multirow}
\usepackage{setspace}
\usepackage{verbatim}
\usepackage{color}
\numberwithin{equation}{section}
\usepackage{hyperref}
\begin{document}

\title{Diffusion constant of supercharge density in $N=4$ SYM at finite chemical potential}
\date{}
\author{Konstantina Kontoudi and Giuseppe Policastro}

\maketitle

\thispagestyle{empty}

\begin{center}
 \emph{Laboratoire de Physique Th\'eorique}\footnote{Unit\'e Mixte du CRNS et
    de l'Ecole Normale Sup\'erieure associ\'ee \`a l'universit\'e Pierre et
    Marie Curie 6, UMR
    8549.}\\
\emph{ Ecole Normale Sup\'erieure,  \\
24 rue Lhomond, F--75231 Paris Cedex 05, France} \\
 \textrm{kontoudi@lpt.ens.fr,  policast@lpt.ens.fr}
\end{center}

\vskip 1cm 

\abstract{We compute holographically the diffusion constant of supercharges in N=4 SYM at finite chemical potential for the R-charge, by solving the equations of motion for the  transverse mode of the gravitino in the  STU black hole in 5 dimensions. We consider the case of one charge and three  charges, and we present analytical solutions for small values of the charges  and numerical solutions for arbitrary values. We compare our results with other known results in 4 dimensions. }

\newpage

\tableofcontents

\setcounter{page}{1}
\setcounter{figure}{0}
\section{Introduction}

A large body of literature has been devoted to the study of the transport and hydrodynamic properties of strongly coupled gauge theories 
using the AdS/CFT correspondence, starting from  \cite{Policastro:2001yc, Policastro:2002se, Policastro:2002tn}; see \cite{CasalderreySolana:2011us} for a recent review. 

One of the most striking results of these investigations has been the discovery that the ratio of the shear viscosity to the entropy density is a universal quantity, that has the same value of $1/4 \pi$ in every theory that admits a description in terms of Einstein gravity 
coupled to matter fields; the universality is spoiled if the Lagrangian has higher-derivative terms, which means generically, given the dictionary of the holographic correspondence, that it holds only in the limit of infinitely strong coupling of the dual field theory. See \cite{Cremonini:2011iq} for a review. 
The underlying reason is the uniqueness of Einstein gravity, or in other words the universal coupling of gravity to all matter fields; 
as a consequence, it can be shown that in any planar black hole background, the transverse graviton (i.e. with polarization parallel 
to the directions in which the horizon extends) satisfies the same equation of motion; the shear viscosity is related to the 
absorption of this mode by the black hole horizon. 

This result was put in a more general framework by \cite{Iqbal:2008by}, who showed that 
for some transport coefficients, the relevant correlator has no radial flow, or it is independent of the radial direction in the low-energy limit; this happens whenever the fluctuating mode that couples to the operator is a massless scalar. In that case, the transport coefficient that is in principle computed on the boundary can also be computed on the horizon, and so typically it has less dependence 
on all the details of the model; but it is not enough to imply universality as there can still be a residual dependence of the 
"effective coupling" at the horizon. 

Most of the work on the subject has been devoted to bosonic correlators, with relatively little attention paid to the fermions.  
In fact, it is natural to wonder whether the constraints imposed by supersymmetry may be sufficient to imply universality 
for the transport of supersymmetry charges. 
With this question in mind, one of the authors (GP) has computed the supercurrent diffusion constant in N=4 SYM  \cite{Policastro:2008cx}. In that paper the result was obtained by looking at poles of the retarded correlator of the supercharge density ; the imaginary part of the pole gives the attenuation coefficient which can be related to the diffusion constant by general hydrodynamic arguments. 
While this was the simplest way to arrive at the result, it was not optimal because it used the longitudinal channel instead of the transverse one; even for the graviton the universality would not be manifest in this channel. 

In the present paper we  reanalyze the question by looking at the transverse mode of the gravitino. First we consider an uncharged gravitino in a generic black-brane background; the technique of radial flow allows us to derive a general formula for the diffusion constant; this formula reproduces the known result in the case of the AdS black hole; however even in this simple case the fluctuation is not massless in general, so we cannot read the value in terms of horizon data. 

Next we consider what happens when the gravitino is charged, as it is the case in gauged supergravity. The equations of motion in this case are much more complex and the radial flow is not useful. Therefore we consider only one particular case: the black hole in the so-called STU model.  This is a solution of N=2 gauged supergravity, but it can be embedded in N=8 and as such corresponds to 
taking $N=4$ SYM at finite chemical potential for the R-charge. The STU black hole can be charged under three commuting $U(1)$ gauge fields. We consider the special cases of one charge and three equal charges. In each case we solve numerically the equations of motion and compute the diffusion constant as a function of the ratio of temperature over chemical potential. 

This is the plan of the paper: in section \ref{intro} we review the properties of the supersymmetry correlators in the hydrdynamic regime, and the setup for the holographic computation; in section \ref{dif} we derive the diffusion constant for the uncharged gravitino; in section \ref{stu} 
we review the solution for the STU black hole and their thermodynamic properties; in sections \ref{exp}-\ref{num} we detail the computation of the 
diffusion constant and give the results (in fig. 1-4). 
The appendices contain some details on the derivation of the Kubo formula for the supercurrents, and the normalization 
of the boundary action.

\section{Supercharge diffusion at zero density}
\subsection{Supercurrent correlator}
\label{intro}
A supersymmetric field theory at finite temperature has hydrodynamic excitations corresponding to fluctuations of the 
supercharge density (see \cite{LebedevSmilga} and  \cite{Kovtun:2003vj} for more details on hydrodynamics with supercharges). 
In particular there is a supersymmetric sound wave that propagates with a speed and attenuation that are different from those of ordinary 
sound. The attenuation rate is governed by the diffusion constant of the supercharge; this has been computed in the N=4 SYM theory 
for the first time in  \cite{Policastro:2008cx} using holography. In this section we compute it in a different way, not using  the sound attenuation but the transverse part of the correlator as explained below; along the way we present the  definitions and techniques that will be used in the next section. 
 
The diffusion constant can be extracted by means of a Kubo formula (see appendix \ref{app_Kubo}) from the retarded correlator of the supercurrents $S_i^{\alpha}$,  defined as:
\begin{equation}
G_{ij}^{\alpha \dot{\beta}}(k)=\int d^4xe^{-ik\cdot x}i\theta(x^0)\langle\{S_i^{\alpha}(x),\bar S_j^{\dot{\beta}}(0)\}\rangle
\end{equation}
Conservation of the supercurrents and superconformal invariance imply:
\begin{equation}
k^iG_{ij}=0\ \ \ \ \gamma^iG_{ij}=G_{ij}\gamma^j=0
\end{equation}
The correlator of the supercurrents can be build out of projectors that implement these constraints. The projector on the transverse gamma-traceless part of a vector-spinor is 
\begin{equation}
P_i^j=\delta_i^j-\frac{1}{3}\left(\gamma_i-\frac{k_i\slashed{k}}{k^2}\right)\gamma^j-\frac{1}{3k^2}(4k_i-\gamma_i\slashed{k})k^j
\end{equation}
The correlator can be written as $G_{ij}=P_i^kM_{kl}P^l_j$. At zero temperature, Lorentz invariance dictates the form of $M$ to be $M_{kl}=A(k^2)\, \slashed{k}\, \eta_{kl}$. At finite temperature or density Lorentz invariance is broken and  the correlator can depend apart from $k$ also on the velocity of the fluid $u^i=(1,0,0,0)$. One can write another projector that is transverse both to $k$ and $u$:  
\begin{equation}
\begin{split}
& P^T_{11}=P^T_{1i}=0\\
&P^T_{jk}=\delta_{jk}-\frac{1}{2}\left(\gamma_j-\frac{q_j\slashed{q}}{q^2}\right)\gamma_k-\frac{1}{2q^2}(3q_j-\gamma_j\slashed{q})q_k
\end{split}
\end{equation}
where $k^i=(\omega-\mu,\mathbf{q})$. We define the longitudinal projector as $P^L=P-P^T$. The correlator can have three possible structures: $P^LP^L$,  $P^TP^T$  and $P^LP^T+P^TP^L$. In the following we choose $\mathbf{q}=(0,0,q)$. With this choice the mixed correlator vanishes identically.  The longitudinal and the transverse part of the correlator can be written as:
\begin{equation}
\begin{split}
& G^L_{ij}=(P^L)_i^k M_{km}(P^L)_{mj}\\
& G^T_{ij}=(P^T)_i^k \tilde M_{km}(P^T)_{mj}
\end{split}
\end{equation}
and the most general  form of $M$ and $\tilde M$ allowed by the symmetries is:
\begin{equation}
\begin{split}
& M_{km}=  \eta_{km} (a \slashed{k} + a' \slashed{u} ) + u_k u_m (b \slashed{k} + b' \slashed{u} )   \\
& \tilde M_{km} = \eta_{km} (\tilde a \slashed{k} + \tilde a' \slashed{u})
\end{split}
\end{equation}
There are additional relations between the coefficients that come from requiring rotational invariance at zero momentum. 
For the following we will need these relations between the components of the  correlator in the low momentum limit:
\begin{equation}
\begin{split}
&G^T_{22}=G^T_{33}=-a'/2\ \ \ \ \ \ \ G^T_{44}=0\\
&G^L_{22}=G^L_{33}=-a'/6\ \ \ \ \ \ \ G^L_{44}=-2a'/3\\
\end{split}
\end{equation}
The longitudinal part of the correlator was calculated in \cite{Policastro:2008cx}.
In the present article we focus on the transverse correlator.

\subsection{Transverse gravitino in blackbrane background}
For the holographic computation we have to consider a gravitino propagating in the near-extremal 3-brane background (see \cite{Corley:1998qg} for the analogous computation at zero temperature). 
The near horizon geometry has the following metric:
\begin{equation}\label{metric1}
ds^2=\frac{\pi^2 T^2R^2}{u}(-f(u)dt^2+dx^2+dy^2+dz^2)+\frac{R^2}{4f(u)u^2}du^2
\end{equation}
where $f(u)=1-u^2$. The boundary is located at $u=0$ and the horizon at $u=1$.\\
The bulk action for the gravitino is:
\begin{equation}
S=\int d^4x\sqrt{-g}(\bar{\Psi}_{\mu}\Gamma^{\mu\nu\rho}D_{\nu}\Psi_{\rho}-m\bar{\Psi}_{\mu}\Gamma^{\mu\nu}\Psi_{\nu})
\end{equation}
The covariant derivative acts on a spinor as:
\begin{equation}
D_{\mu}=\partial_{\mu}+\frac{1}{4}\omega^{ab}_{\mu}\gamma_{ab}
\end{equation}
and the equation of motion is the Rarita-Schwinger equation:
\begin{equation}
\Gamma^{\mu\nu\rho}D_{\nu}\Psi_{\rho}-m\Gamma^{\mu\nu}\Psi_{\nu}=0
\end{equation}
In the gauge $\Gamma^\mu \Psi_\mu=0$  the equations for the spatial components of the field read  \cite{Policastro:2008cx}:
\begin{equation}\label{eomi}
\gamma^{5}\psi^{\prime}_k+\frac{1}{2\pi T\sqrt{uf}}\left(\frac{1}{\sqrt{f}}\gamma^1\partial_t+\gamma^j\partial_j\right)\psi_k+\frac{u^2-2}{2uf}\gamma^5\psi_k-\frac{1}{2u}\gamma_k\psi_5+\frac{mR}{2u\sqrt{f}}\psi_k=0
\end{equation}
where $\psi_a=e^{\mu}_a\Psi_{\mu}$ , $\gamma_a=e^{\mu}_a\Gamma_{\mu}$ and $\psi^{\prime}=\partial\psi/\partial u$.\\
Define the Fourier transform as:
\begin{equation}
\Psi_{\mu}(u,x)=\frac{1}{(2\pi)^2}\int d^4ke^{ik_{\nu}x^{\nu}}\Psi_{\mu}(u,k)
\end{equation}
with $k^{\mu}=2\pi T(\omega,0,0,q)$. \\
The field $\eta=\gamma^2\psi_2-\gamma^3\psi_3$ has spin 3/2  under the $O(2)$ rotational symmetry that preserves $k^\mu$ and therefore it decouples from the other components that have spin 1/2. It is easy to see that this is the field that is dual to the transverse 
part of the correlator. Multiplying (\ref{eomi}) for $k=2$ by $\gamma_2$ and for $k=3$ by $\gamma_3$ and subtracting we arrive at the equation satisfied by the transverse component of the gravitino:
\begin{equation}\label{eta_eq}
\eta^{\prime}=-\gamma^5\left(\frac{\slashed{P}}{\sqrt{uf}}+\frac{u^2-2}{2uf}\gamma^5-\frac{3}{4u\sqrt{f}}\right)\eta
\end{equation}
with $\slashed{P}=-i\omega/\sqrt{f}\gamma^1+iq\gamma^4$.\\ 
We cannot solve equation (\ref{eta_eq}) analytically but we can solve it as an expansion in momenta. 
A useful remark is that equation (\ref{eta_eq}) commutes with $\gamma_{23}=diag(1,-1,1,-1)$ and therefore we can project on its eigenvectors. In the eigenspace $\gamma_{23}=1$ (resp. $-1$)  equation (\ref{eta_eq}) becomes a matrix differential equation for $(\eta_1,\eta_3)$ (resp. $(\eta_2,\eta_4)$). 

We find the following solutions near the horizon :
\begin{equation}
\begin{split}
& (1-u)^{-\frac{1}{4}-i\frac{\omega}{2}}(-i,1)\\
& (1-u)^{-\frac{1}{4}+i\frac{\omega}{2}}(i,1)
\end{split}
\end{equation}
We have to impose incoming boundary conditions at the horizon in order to calculate the retarded correlator \cite{Son:2002sd}.We can keep the incoming waves at the horizon by imposing the relation $\eta^+=-i\eta^-$. For the two other  components $(\eta_2,\eta_4)$ the equations are the same after the exchange $q\rightarrow -q$. Repeating the analysis for these components we find the solution up to first order in momenta.
The near boundary expansion of this solution reads:
\begin{equation}
\eta=\left(\begin{array}{l}
u^{3/4}(-i3^{3/4})(q-\omega)\alpha+u^{7/4}\frac{3^{3/4}}{2\sqrt{2}}\beta(\omega,q)\\
u^{3/4}i3^{3/4}(\omega+q)\gamma+u^{7/4}\frac{3^{3/4}}{2\sqrt{2}}\delta(\omega,q)\\
u^{1/4}3^{3/4}[i\beta(\omega,q)+\sqrt{2}(q-(1-\sqrt{2}\slashed{L})\omega)\alpha]\\
u^{1/4}3^{3/4}[i\delta(\omega,q)-\sqrt{2}(q+(1-\sqrt{2}\slashed{L})\omega)\gamma]
\end{array}
\right)
\end{equation}
where $\slashed{L}=\log(1+\sqrt{2})$.
We may identify the source of the field $\varphi$ with the negative chirality part $\eta^{-}=u^{1/4}\varphi$ and identify $\chi$ and $\tilde{\chi}$ as the coefficients of the positive chirality expansion $\eta^{+}=u^{3/4}\tilde{\chi}+u^{7/4}\chi$: 
\begin{equation}
\varphi=
\left(\begin{array}{l}
3^{3/4}[i\beta(\omega,q))+\sqrt{2}(q-(1-\sqrt{2}\slashed{L})\omega)\alpha]\\
3^{3/4}[i\delta(\omega,q)-\sqrt{2}(q+(1-\sqrt{2}\slashed{L})\omega)\gamma]
\end{array}\right)
\end{equation}

\begin{equation}
\chi=
\left(\begin{array}{l}
\frac{3^{3/4}}{2\sqrt{2}}\beta(\omega,q)\\
\frac{3^{3/4}}{2\sqrt{2}}\delta(\omega,q)
\end{array}\right) \quad \quad \tilde{\chi}=
                                    \left(\begin{array}{c}
                                   i3^{3/4}(\omega-q)\alpha\\
                                   i3^{3/4}(\omega+q)\gamma
                                   \end{array}\right)
\end{equation}

Multiplying $\phi$ by $\slashed{k}=-i2\pi T\operatorname{diag}(\omega-q,\omega+q)$ and neglecting terms that are of higher order in the momenta we can express   $\tilde{\chi}$ in terms of the source:
\begin{equation}
\tilde{\chi}=\frac{i}{2\pi T}\slashed{k}\varphi
\end{equation}
Similarly we can verify that the following expression is correct up to higher order terms in the momenta:
\begin{equation}
\varphi=\operatorname{diag} (2\sqrt{2}i+4q-4(1-\sqrt{2}\slashed{L} )\omega,2\sqrt{2}i-4q-4(1-\sqrt{2}\slashed{L} )\omega)\chi
\end{equation} 
Inverting the matrix and expanding on small momenta we can express   $\chi$ in terms of the source:
\begin{equation}
\chi=\operatorname{diag} (-i\frac{\sqrt{2}}{4}-4q+4(1-\sqrt{2}\slashed{L} )\omega,-i\frac{\sqrt{2}}{4}+4q+4(1-\sqrt{2}\slashed{L} )\omega)\varphi
\end{equation}

\subsection{Computation of the diffusion constant}
\label{dif}
Given that, for the Rarita-Schwinger field, the on-shell action vanishes , the only contribution to the correlator will come from the boundary action:
\begin{equation}
S_{bdy}=\mathcal{N}\int d^4x \sqrt{-h}h^{ij}\bar{\Psi}_i\Psi^j
\end{equation}
where $h$ is the induced metric on the boundary and $\bar{\Psi}=i\Psi^{\dagger}\Gamma^1$.  Performing the field redefinitions $\phi=\gamma^2\psi_2+\gamma^3\psi_3$ and $\eta=\gamma^2\psi_2-\gamma^3\psi_3$ and setting $\phi=0$, $\psi_1=0$, $\psi_4=0$ since we are only interested in the transverse part we get:
\begin{equation}
S_{bdy}=\frac{\mathcal{N}}{2}\int d^4x\sqrt{-h}\frac{\sqrt{u}}{\sqrt{f}\pi TR}\bar{\eta}\eta
\end{equation}
Fourier transforming and taking the limit $u=\epsilon\rightarrow 0$ we have:
\begin{equation}
S_{bdy}=\frac{\mathcal{N}}{2}\frac{\sqrt{\epsilon}}{\pi TR}\int d^4k \sqrt{-h}  \left[\eta^\dagger_-(k)\eta_+(k)+\eta^\dagger_+(k)\eta_-(k)\right]
\end{equation}
The correlator that we want to calculate is not hermitian and therefore we should keep only the first term in the action to extract the correlator. This procedure is equivalent to the recipe proposed in \cite{Iqbal:2009fd}. Substituting the expansion of the field on the boundary the action becomes:
\begin{equation}
S_{bdy}=\frac{\mathcal{N}}{2}(\pi TR)^3\int d^4k\left(\epsilon^{-1/2}\varphi^{\dagger}\tilde{\chi}+\epsilon^{1/2}\varphi^{\dagger}\chi\right)
\end{equation}
To substract the divergencies in the limit $\epsilon\rightarrow 0$ we introduce the counterterm:
\begin{equation}
S_{ct}=-S^{div}_{bdy}=-\frac{\mathcal{N}}{2}(\pi TR)^3\int d^4k\epsilon^{-1/2}\varphi^{\dagger}\tilde{\chi}
\end{equation}
The counterterm is written in covariant form as:
\begin{equation}
S_{ct}=-\frac{\mathcal{N}R}{4}\int d^4x \sqrt{-h}\bar{\eta}_-\slashed{\partial}_h\eta_-
\end{equation}
where $\bar{\eta}_-=\eta^{\dagger}\Gamma^1$ and $\slashed{\partial}_h=\Gamma^{\mu}\partial_{\mu}$. This counterterm does not introduce any finite correction to the correlator since there is no $u^{7/4}$ term in the boundary expansion of $\eta_-$. The correlator is calculated from the renormalized action.
\begin{gather*}
 S_{ren}=S_{bdy}+S_{ct}=\\
 \frac{\mathcal{N}}{2}(\pi TR)^3\int\varphi^{\dagger}\operatorname{diag} (-i\frac{\sqrt{2}}{4}-4q+4(1-\sqrt{2}\slashed{L} )\omega,-i\frac{\sqrt{2}}{4}+4q+4(1-\sqrt{2}\slashed{L} )\omega)\varphi
\end{gather*}
The retarded correlator of the operator dual to $\eta$ is therefore:
\begin{equation}
 G^{\mathcal{R}}=\frac{\mathcal{N}}{2}(\pi TR)^3\operatorname{diag} (-i\frac{\sqrt{2}}{4}-4q+4(1-\sqrt{2}\slashed{L} )\omega,-i\frac{\sqrt{2}}{4}+4q+4(1-\sqrt{2}\slashed{L} )\omega)
\end{equation}
where we have absorbed the extra $\epsilon$ power in a field redefinition. The normalization of the action is calculated in the appendix and is found to be:
\begin{equation}
 \mathcal{N}=\frac{N_c^2}{\pi^2}
\end{equation}
From the analysis of the projections of the correlators we can deduce that with i indices summed $G^{ii}_{\alpha\dot{\alpha}}=4G^{\mathcal{R}}_{\alpha\dot{\alpha}}$ where $G^{ii}_{\alpha\dot{\alpha}}$ is the correlator that comes in the Kubo formula (\ref{Kubo}) for the diffusion constant. The energy density is:
\begin{equation*}
 \varepsilon=\frac{3\pi^2}{8}N_c^2T^4
\end{equation*}
We finally compute the diffusion constant from the Kubo formula and find:
\begin{equation}\label{Dzero}
 D_s=\frac{2\sqrt{2}}{9\pi T}
\end{equation}
This result is in agreement with the diffusion constant calculated in \cite{Policastro:2008cx}.
\subsection{Running of the correlator}

In this section we give an alternative derivation in the spirit of the  membrane paradigm approach  to the question of universality of the diffusion constant \cite{Iqbal:2008by}. The idea is that the correlator can be formally computed on any radial slice of the geometry. 
When evaluated on the boundary it corresponds to the real field-theory correlator, and one can derive a flow equation that expresses the running from the boundary to the horizon. In favorable cases, at low frequency the flow is trivial and the UV correlator is equal to the value at the horizon. 

Let us consider again the Rarita-Schwinger equation (\ref{eta_eq}); at zero momentum and frequency, it becomes 
\begin{equation}
 \eta_+' = (-A + B) \eta_+ \,, \quad \eta_-' = (-A - B) \eta_- 
\end{equation}
where $B = m \sqrt{g_{uu}}$, and $m=3/2$ for the gravitino.  The ratio $w = \eta_+ / \eta_-$ satisfies $\partial_u w = 2 B w$, and we have the flow equation 
\begin{equation}
{\cal F}' \equiv  \partial_u \left( e^{-2 \int B(u) du} \, {\eta_+ \over \eta_-} \right) = 0 \,. 
\end{equation}
The integral converges at the horizon, so we have ${\cal F}(u=1) = \eta_+/\eta_-(u=1) = - i$. At the boundary we have ${\cal F} \sim u^{-m} \eta_+ / \eta_- = \chi / \phi$, 
but the coefficient of proportionality requires the computation of $\int_1^u \sqrt{g_{uu}}$ so we cannot express it purely in terms of horizon data.  We find 
\begin{equation}
 D_s = {8 \over 9 \pi T} exp \left( -m \log \epsilon - 2 m \int_\epsilon^1 \sqrt{g_{uu}} du \right)
\end{equation}
 This formula reproduces the result (\ref{Dzero}), and allows to compute the diffusion constant in any asymptotically AdS black hole background.  
 However, as we will see, in presence of a finite charge density for a $U(1)$ under which the gravitino is charged (typically an R-symmetry), the equations of motion are more complicated and do not become diagonal even at zero momentum.

\section{Supercharge diffusion at finite density} 
\subsection{The background}
\label{stu}

We want to consider the change in the diffusion constant when there is a finite background charge density. In N=4 SYM there is an $SU(4)$ R-symmetry group under which the supercurrents transform in the fundamental representation. It is possible to give 
a vev simultaneously to three charges in the Cartan subalgebra of $SU(4)$. The corresponding solution  is an asymptotically AdS black hole carrying electric charges, known as the STU black hole. It was first constructed as a solution of  $d=5,\mathcal{N}=2$ gauged supergravity with 2 matter multiplets. The construction of the theory is explained in \cite{Gunaydin:1984ak} and the solutions of the bosonic part are presented in \cite{Behrndt:1998jd} . The solution can be embedded in type IIB supergravity, the uplift to 10 dimensions is given in \cite{Cvetic:1999xp} but for our purposes it is sufficient to work with the 5-dimensional solution. \\
The theory contains three abelian gaueg fields $A_I$ and three scalar fields subject to the constraint $X^1X^2X^3=1$ , with the metric on the scalar manifold  given by:
\begin{equation}
 G^{IJ}=2\operatorname{diag}[(X^1)^2,(X^2)^2,(X^3)^2]
\end{equation}
and the scalar potential is $V = 2 \sum_I 1/X^I$. 
The metric of the black hole with translationally invariant horizon is:
\begin{gather*}
 ds^2=-\mathcal{H}^{-2/3}\frac{(\pi T L)^2}{u}f(u)dt^2+\mathcal{H}^{1/3}\frac{(\pi T L)^2}{u}(dx^2+dy^2+dz^2)+\mathcal{H}^{1/3}\frac{L^2}{4f(u)u^2}du^2\\
f(u)=\mathcal{H}(u)-u^2\prod_{I=1}^3(1+\kappa_I)\ \ \ \ \ H_I(u)=1+\kappa_Iu\\
 \kappa_I=\frac{q_I}{(\pi T L^2)^2}\ \ \ \ \ \mathcal{H}(u)=\prod_{I=1}^3H_I(u)
\end{gather*}
where $q_I$ are the three charges. In the zero charge limit this metric is exactly (\ref{metric1}). The scalar and gauge fields are given by:
\begin{equation}
 X^I=\frac{\mathcal{H}^{1/3}}{H_I(u)}\ \ \ \ \ A_t^I=\pi T L u\frac{\kappa_I}{H_I(u)}\prod_{J=1}^3(1+\kappa_J)
\end{equation}
The Hawking temperature of the background is:
\begin{equation}
T_H=\frac{2+\kappa_1+\kappa_2+\kappa_3-\kappa_1\kappa_2\kappa_3}{2\sqrt{(1+\kappa_1)(1+\kappa_2)(1+\kappa_3)}}T
\label{temp}
\end{equation} 
The energy density, entropy density, charge density and chemical potentials are respectively:
\begin{equation}
\begin{gathered}
 \epsilon=\frac{3\pi^2N^2T^4}{8}\prod_{I=1}^3(1+\kappa_I)\\
 s=\frac{\pi^2N^2T^3}{2}\prod_{I=1}^3(1+\kappa_I)^{1/2}\\
 \rho_I=\frac{\pi N^2T^3}{8}\sqrt{\kappa_I}\prod_{J=1}^3(1+\kappa_J)^{1/2} \\
 \mu_I=A^I_t(u)|_{u=1}=\frac{\pi TL\sqrt{\kappa_I}}{1+\kappa_I}\prod_{J=1}^{3}(1+\kappa_J)^{1/2}
\end{gathered}
\end{equation}
Imposing the condition of thermodynamic stability on the background implies a restriction on the charges of the background \cite{Son:2006em}
\begin{equation}
2-\kappa_1-\kappa_2-\kappa_3+\kappa_1\kappa_2\kappa_3>0
\label{thermo}
\end{equation}
The relevant part of the Lagrangian for the gravitino is \cite{Gunaydin:1984ak}:
\begin{equation}
 \begin{split}
  \mathcal{L}=& e\left[\right.\bar{\Psi}_{\mu}\Gamma^{\mu\nu\rho}\mathcal{D}_{\nu}\Psi_{\rho}+i\bar{\lambda}^a\Gamma^{\mu}\Gamma^{\nu}\Psi_{\mu}f^a_i\partial_{\nu}\phi^i-\frac{1}{2}h_I^a\bar{\lambda}^a\Gamma^{\mu}\Gamma^{\lambda\rho}\Psi_{\mu}F^I_{\lambda\rho}\\
& +\frac{1}{8X^I}i\left(\bar{\Psi}_{\mu}\Gamma^{\mu\nu\rho\sigma}\Psi_{\nu}F^I_{\rho\sigma}+2\bar{\Psi}^{\mu}\Psi^{\nu}F^I_{\mu\nu}\right)+\frac{3}{2L}\bar{\Psi}_{\mu}\Gamma^{\mu\nu}\Psi_{\nu}\mathcal{V}_o-\frac{3}{L}\bar{\lambda}^a\Gamma^{\mu}\Psi_{\mu}\mathcal{V}_a\left.\right]
 \end{split}
\end{equation}
where $I=1,2,3$, $i,a=1,2$. $h_I^a$, $\mathcal{V}_a$ and $\mathcal{V}_o$ are functions of the scalar fields , $\phi^i$ are the scalars fields on the constrained scalar manifold and $f^a_i$ is the vielbein on this manifold. The covariant derivative acts on a spinor as
\[\mathcal{D}_{\mu}\psi=D_{\mu}\psi-i\frac{3}{2L}V_IA^I_{\mu}\psi\]
where $V_I$ are constants and should be equal to $1/3$ for the $H_I$ to be properly normalized \cite{Behrndt:1998jd}. $\mathcal{V}_o$ is given by:
\begin{equation}
 \mathcal{V}_o=\frac{1}{2}G^{IJ}\frac{V_I}{X^J}
\end{equation}
and $F^I_{\mu\nu}=\partial_{\mu}A^I_{\nu}-\partial_{\nu}A^I_{\mu}$ are the gauge field strengths.The equations of motion for the gravitino are:
\begin{equation}
\begin{split}
\Gamma^{\mu\nu\rho}\mathcal{D}_{\nu}\Psi_{\rho}+  i\Gamma^{\mu}\Gamma^{\nu}&\lambda_af^a_i\partial_{\nu}\phi^i-\frac{1}{2}h_I^a\Gamma^{\mu}\Gamma^{\lambda\rho}\lambda^aF^I_{\lambda\rho} +\frac{i}{8X^I}\left(\Gamma^{\mu\nu\rho\sigma}\Psi_{\nu}F^I_{\rho\sigma}+2\Psi^{\nu}F^I_{\mu\nu}\right)\\
&+\frac{3}{2L}\Gamma^{\mu\nu}\Psi_{\nu}\mathcal{V}_o-\frac{3}{L}\Gamma^{\mu}\lambda^a\mathcal{V}_a=0
\end{split}
\end{equation}
The supersymmetry transformations act on the gravitino as:
\begin{equation}
\delta\Psi_{\mu}=  \mathcal D_{\mu}\epsilon+\frac{i}{24X^I}(\Gamma^{\nu\rho}_{mu}-4\delta^{\nu}_{\mu}\Gamma^{\rho})\hat F^I_{\nu\rho}\epsilon+\frac{1}{6L}\sum_IX^I\Gamma_{\mu}\epsilon
\end{equation}
where
\[\hat F^I_{\mu\nu}= F^I_{\mu\nu}+\frac{i}{4X^I}\bar\Psi_{[\mu}\Psi_{\nu ]}\]
and we neglected the terms containing the spinors $\lambda$ in the two previous formulas.\\
We are again interested in the transverse component of the gravitino, which can be defined as $\eta=\Gamma^x\Psi_x-\Gamma^y\Psi_y$. Using the same strategy as in the zero-charge case, one can show that the transverse part of the gravitino is decoupled, both from the longitudinal part and from the spinors $\lambda$.  The equation for the transverse component is therefore:
\begin{equation}\label{chargedgraveom}
\eta'+\left(\frac{\gamma^5\slashed{K}}{\sqrt{fu}}+F(u)-\frac{3i}{2L}\sqrt{g_{uu}g^{tt}}\gamma_5\gamma^1V_IA^I_t+\frac{i}{4X^I}\sqrt{g^{tt}}\gamma^1F^I_{ut}+\frac{3}{2L}\sqrt{g_{uu}}\gamma^5\mathcal{V}_o\right)\eta=0
\end{equation}
with
\begin{equation}
\begin{split}
&\slashed{K}=-i\sqrt{\frac{\mathcal H}{f}}\gamma^1\omega+iq\gamma^4\\
& F(u)=\frac{1}{4}\frac{g'_{tt}}{g_{tt}}+\frac{3}{4}\frac{g'_{xx}}{g_{xx}}
\end{split}
\end{equation}
Note that the transverse component of the gravitino is gauge invariant and that we can again split our equation in two different systems by projecting on the eigenspaces of  $\gamma_{23}$.
\subsection{Analytic solution for small charge}	    
\label{exp}
The system of differential equations obtained above cannot be solved analytically. Since calculating the diffusion constant requires knowing the retarded correlator for zero momenta we set the momentum to zero and attempt to find a solution as an expansion in the charges. We consider in turn the cases where only one charge is turned on, and when there are three equal charges. \\

\subsubsection*{One charge}
In this case we take  $\kappa_1=\kappa,\  \kappa_2=\kappa_3=0$ and attempt to find a solution of the form:
\begin{equation}
\eta(u,\kappa)=\eta^{(0)}(u)+\sqrt{\kappa}\ \eta^{(1)}(u)+\kappa\ \eta^{(2)}(u)+…
\label{kappa_exp}
\end{equation}
Solving the system of equations near the horizon we identify the exponents:
\begin{equation}
U_{\pm}=-\frac{1}{4}\pm\frac{i}{4}\frac{\sqrt{\kappa}+4\sqrt{1+\kappa}\ \omega}{2+\kappa}
\end{equation}
The choice of $U_-$ yields incoming-wave boundary conditions. 
Near the boundary we find the following behavior:
\begin{equation}
\begin{split}
& \eta_1=\omega\phi(\omega,\kappa)u^{3/4}+ \chi(\omega,\kappa)u^{7/4}+\phi(\omega,\kappa)\frac{\sqrt{\kappa(1+\kappa)}+2\omega^3}{2}u^{7/4}\log u+…\\
& \eta_3=\phi (\omega,\kappa)u^{1/4}+\tilde\phi (\omega,\kappa)u^{5/4}+…
\end{split}
\end{equation}
This expansion implies that there will be a logarithmic divergence of the correlator that would need to be regularized with an appropriate   counterterm. However this will not affect the diffusion constant that is computed from the imaginary part of the correlator, so we will ignore this divergence. 
\par
The integration constants that appear at every order in the $\sqrt\kappa$ expansion (\ref{kappa_exp})  can be fixed at zeroth order by imposing incoming boundary conditions at the horizon and at the i-th order by imposing $\eta^{(i)}=0$ on the horizon. The near boundary expansion up to order $\sqrt{\kappa}$ for the two components is\footnote{For q=0 the system for $(\eta_1,\eta_3)$ is identical to the system for $(\eta_2,\eta_4)$}:
\begin{equation}
\begin{gathered}
 \eta_1=u^{7/4}[2^{-3/4}i+\sqrt{\kappa}(\alpha+\beta\log u)]\\
 \eta_3=u^{1/4}(2^{3/4}-i\sqrt{\kappa}\gamma)\\
\end{gathered}
 \label{sol1}
\end{equation}
where  $\alpha,\beta,\gamma$ are {\it  real}  constants. Notice that there is no term going like $u^{3/4}$ so we don't need to regularize the boundary action. In terms of the source $\phi$ and the normalizable mode $\chi$ the correlator is then simply given by:
\begin{equation}
G^{\mathcal R}=\frac{\pi N_c^2T^3}{2}\frac{\chi}{\phi}=\frac{-\sqrt{2}}{4}i+\left(\frac{\alpha}{2^{3/4}}-\frac{\gamma}{4\ 2^{3/4}}+\frac{\beta}{2^{3/4}}\log u\right) \sqrt{\kappa}+...
\end{equation}
The coefficient of $\sqrt\kappa$ is real and therefore the diffusion constant is not corrected to the order $\sqrt{\kappa}$. 

\subsubsection*{Three charges}
\par
We consider here the case of three equal charges, namely $\kappa_1=\kappa_2=\kappa_3=\kappa$. For this special choice of the charges the thermodynamic stability condition (\ref{thermo}) is saturated for $\kappa=2$ but it is not violated even for $\kappa>2$. The Hawking temperature (\ref{temp}) on the other hand becomes negative for $\kappa>2$ and so in order to have a physical theory we must restrict the value of the charge to $\kappa<2$. For $\kappa=2$ we have an extremal black hole. \\
We attempt to find a solution as an expansion in $\sqrt{\kappa}$ as in the case of one charge. The indices near the horizon are
\begin{equation}
V_{\pm}=-\frac{1}{4}\pm i \frac{3\sqrt{ \kappa(1+\kappa)}+4\omega}{4(2-\kappa)\sqrt{1+\kappa}}
\end{equation}
and the incoming wave boundary condition corresponds to $V_-$. The near boundary expansion is written in this case:
\begin{equation}
\begin{split}
& \eta_1=\omega\phi(\omega,\kappa)u^{3/4}+ \chi(\omega,\kappa)u^{7/4}+\phi(\omega,\kappa)\frac{3\sqrt{\kappa(1+\kappa)}-8\omega^3}{8}u^{7/4}\log u+…\\
& \eta_3=\phi (\omega,\kappa)u^{1/4}+\tilde\phi (\omega,\kappa)u^{5/4}+…
\end{split}
\end{equation}
The calculation done for the one charged case goes through for the three charges case and the end result is similar to (\ref{sol1}) with different numerical values for the constants $\alpha,\beta,\gamma$. In this case too, then, the diffusion constant does not receive 
corrections to order $\sqrt{\kappa}$. \\
The explicit computation of higher orders in this expansion becomes too complicated, therefore we revert to a numerical analysis.

\subsection{Numerical Solution}
\label{num}
 
In order to solve numerically the system of equations we need to provide initial conditions in form of a near horizon expansion where we have imposed the incoming boundary conditions. We will work with the components $(\eta_1,\eta_3)$ since the equations for the other two are exactly identical for $q=0$. We solve the system near the horizon and find a expansion of the form:
\begin{equation}
\eta_{hor}(\omega,u,\kappa)=\eta_{hor}^{(0)}(\omega,\kappa)+\sqrt{1-u}\ \eta_{hor}^{(1)}(\omega,\kappa)+…
\end{equation}
Using the expansion above as an initial condition we integrate the system of equations from the horizon to the boundary.  Then we can extract the correlator from the near boundary region of the solution.\\
 
\begin{figure}[H]
 \center
  \includegraphics[width=10cm]{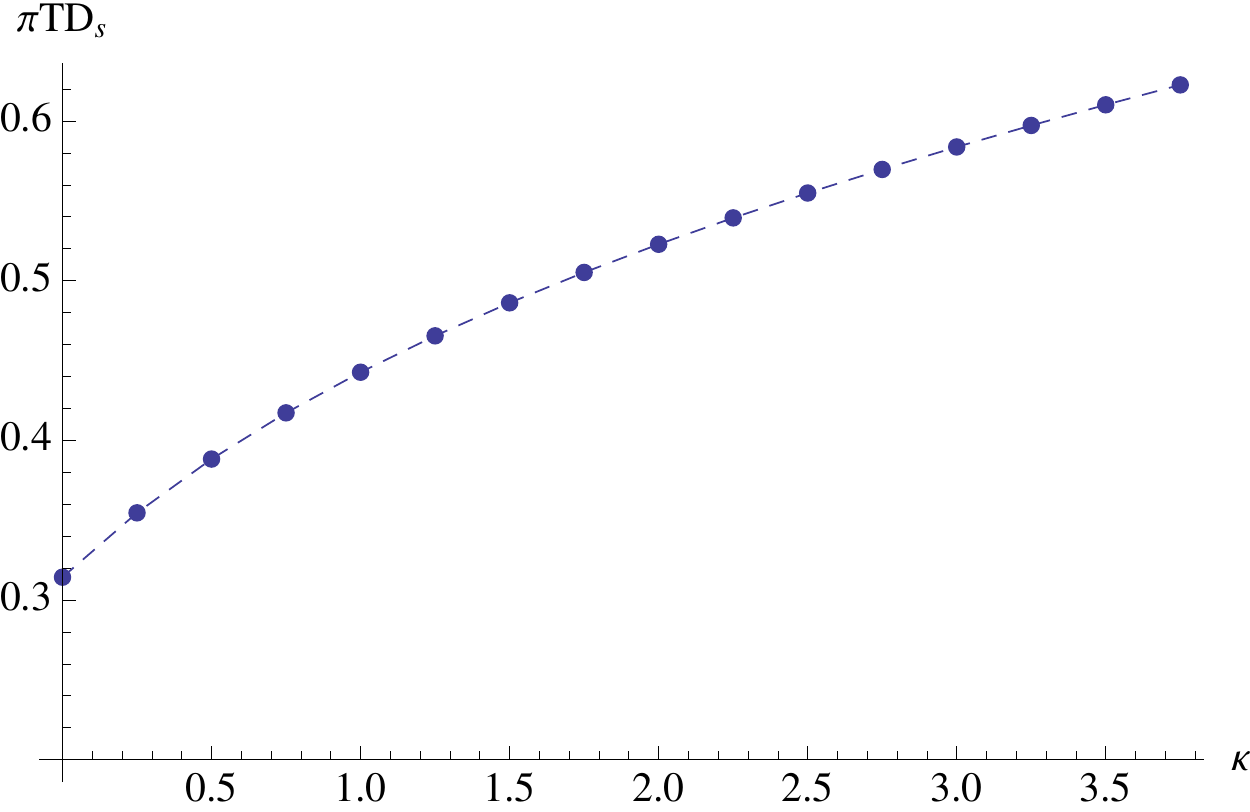}
  \caption{The diffusion constant as a function of the charge for the one charge case $\kappa_1=\kappa,\  \kappa_2=\kappa_3=0$ . }
  \label{1-charge-k}
 \end{figure}

\begin{figure}[H]
 \center
  \includegraphics[width=10cm]{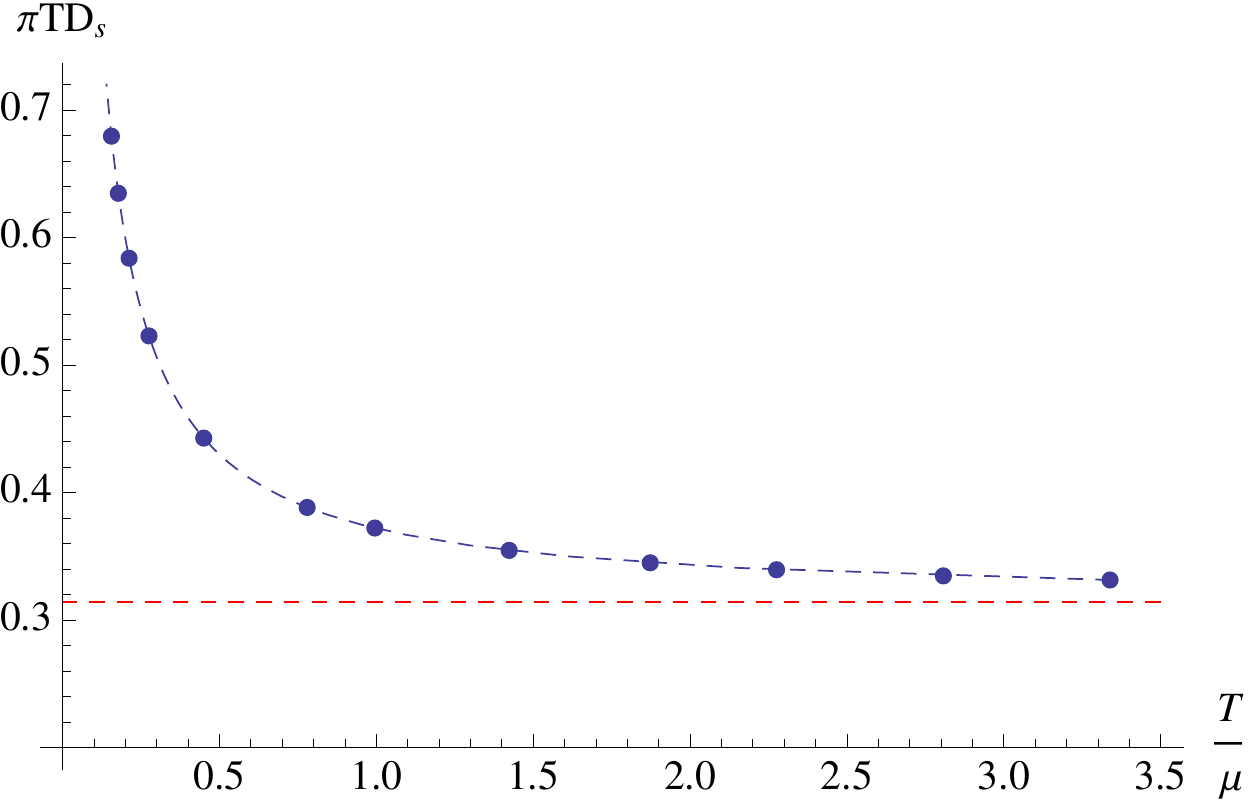}
  \caption{The diffusion constant as a function of $T/\mu$ for the one charge case. The horizontal line represents the zero density analytical result.}
  \label{1-charge-Tmu}
 \end{figure}

 The result for the one-charge case  is presented in figure \ref{1-charge-k}. At $\kappa=0$ it reproduces correctly the value calculated analytically for zero charge. 
In figure \ref{1-charge-Tmu} we present the diffusion constant as a function of $T/\mu$ and find a behavior similar to the one observed in \cite{Gauntlett:2011wm} for an AdS-RN solution of N=2, D=4 supergravity. For large values of $T/\mu$ the diffusion constant approaches the zero density value which is represented by the horizontal line.

\begin{figure}[H]
 \center
  \includegraphics[width=10cm]{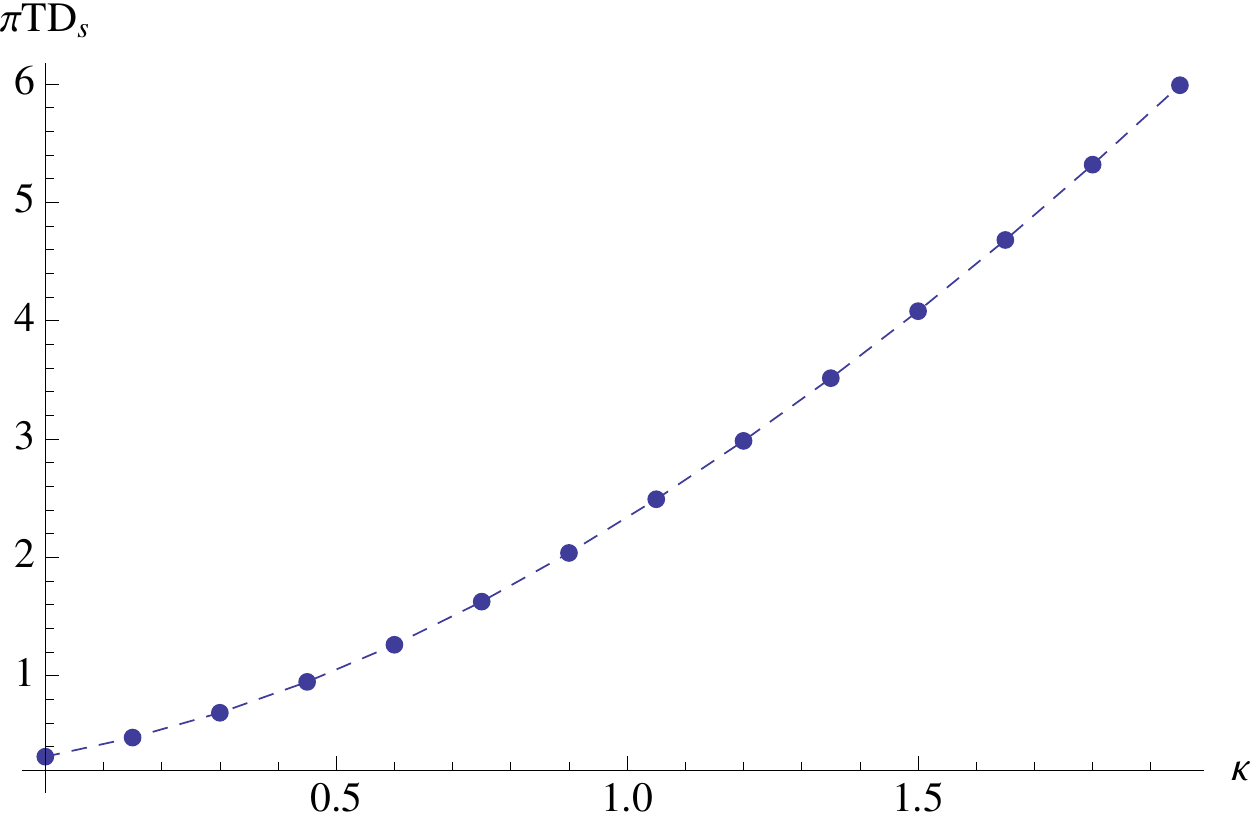}
  \caption{The diffusion constant in the three equal charges case as a function of the charge.}
  \label{3-charge-k}
 \end{figure}

\begin{figure}[H]
 \center
  \includegraphics[width=10cm]{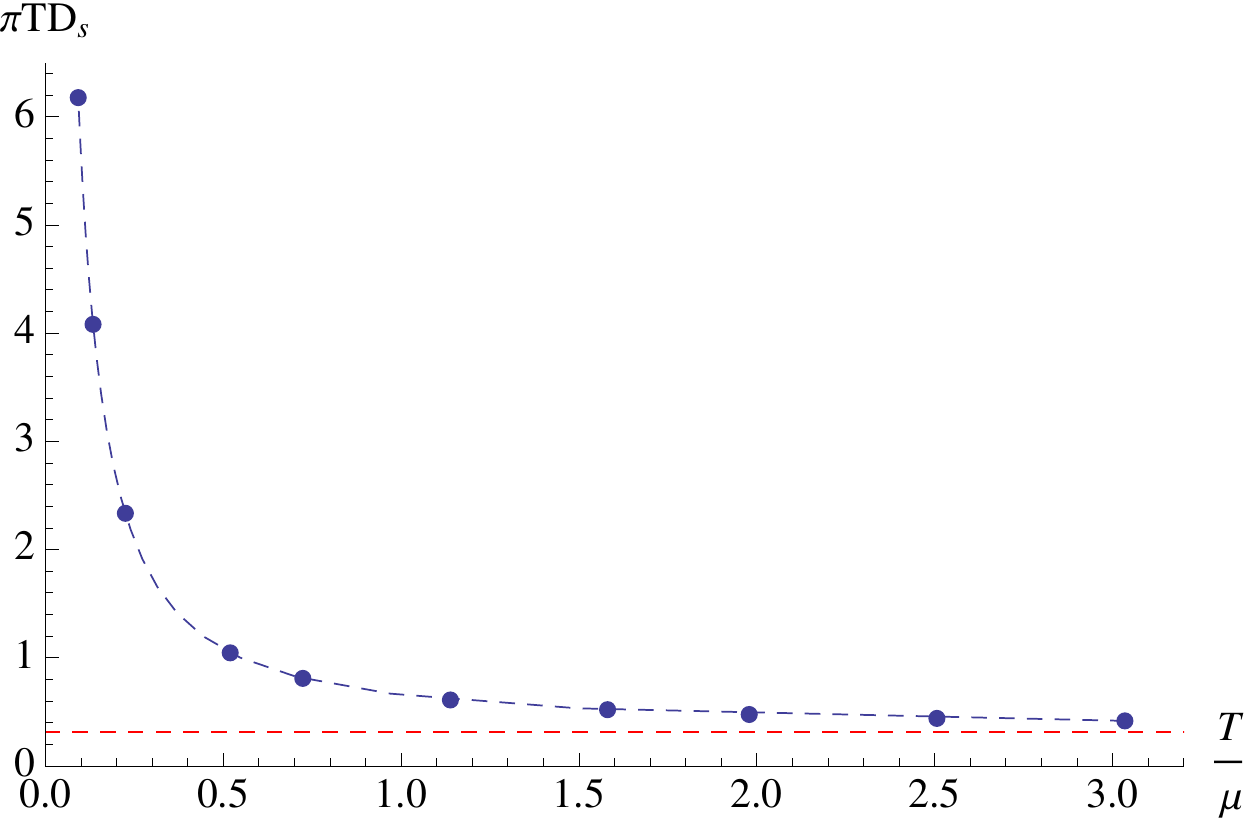}
  \caption{The diffusion constant as a function of $T/\mu$ for the three charges case. The horizontal line represents the zero density analytical result.}
  \label{3-charge-Tmu}
 \end{figure}

In figure \ref{3-charge-k} we give the result for three charges as a function of $\kappa$, and in figure \ref{3-charge-Tmu} as function of 
$T/\mu$. We observe that the behavior is qualitatively similar to that of the one charge case for high temperature, but the diffusion constant (or rather the combination $T D_s$) approaches a finite limit as the temperature goes to zero.  

\section{Conclusions}

We have considered the hydrodynamic limit of the correlators of supercharges in N=4 SYM at zero and finite 
R-charge density. From the transverse part of the correlator the supercharge diffusion constant can be extracted 
using a Kubo formula. In the zero charge case we could confirm the result found previously from the calculation of 
the supersymmetric sound attenuation, using the longitudinal channel of the correlator. 

The main motivation for the present work has been to answer the question of the presence of universality in the 
hydrodynamic supersymmetric sector of theories with a holographic dual. The conclusion appears to be negative: 
the diffusion constant depends on the value of the $U(1)$ charge density.  We have not been able to establish the precise 
form of the dependence, in particular there does not seem to be a simple relation between the diffusion constant and the thermodynamic quantities.  However qualitatively we found a behavior very similar to the one found in the 4-dimensional AdS-RN 
black hole. \\
The absence of universality is perhaps not surprising, since if we consider the form of the  equation (\ref{chargedgraveom})
we see that the transverse gravitino is coupled to the scalars and gauge fields that are running in the solution. \\
We observed that the dimensionless diffusion constant $T D_s$ tends to a finite value for the extremal black hole. It would be interesting to know if 
this is true in all cases of extremal black hole, and perhaps find if this value can be predicted by some sort of attractor mechanism. 

\section*{Acknowledgments}
We would like to thank A. Starinets and particularly P. Kovtun for useful discussions on the issues discussed in appendix \ref{app_Kubo}.

\appendix
\section{Kubo formula derivation}\label{app_Kubo}
The constitutive relation  \cite{Kovtun:2003vj}  which relates the density of charge to the spatial components of the supercurrent is\footnote{When we write $\gamma^{\mu}_{\alpha\dot\alpha}$ we mean the upper right block of the $\gamma$ matrix. We made the replacement $\sigma^{0}\rightarrow i\gamma^1$ and $\sigma^{i}\rightarrow -i\gamma^{i}$ with respect to \cite{Kovtun:2003vj}.}:
\begin{equation}
S^i_{\alpha}=-D_s\partial^i\rho_{\alpha}+D_{\sigma}(\gamma^{ij}\partial_j\rho)_{\alpha}-\frac{P}{\varepsilon}(\gamma^i\gamma^1\rho)_{\alpha}
\end{equation}
Combining with the continuity equation $\partial_t\rho_{\alpha}+\partial_iS^i_{\alpha}$ and Fourier transforming in space yields:
\begin{equation}
[(\partial_t+D_s k^2)\delta_{\alpha}^{\beta}-ic_s(k^j\gamma_j\gamma^1)_{\alpha}^{\beta}]\rho_{\beta}(t,\vec{k})=0
\end{equation}
where $c_s=P/\varepsilon$. The solution of this equation in the limit of small $\vec{k}$ is:
\begin{equation}
\rho_{\alpha}(t,\vec{k})=e^{-D_sk^2t}[\delta_{\alpha}^{\beta}\cos(kc_st)+i(\hat{k}^j\gamma_j\gamma^1)_{\alpha}^{\beta}\sin(kc_st) ]\rho_{\beta}(0,\vec{k})
\end{equation}
From this solution we can deduce the following supercharge density correlators in the small $\vec k$ limit since $C_{\alpha\dot\alpha}(0,\vec 0)$ is fixed by the supersymmetry algebra. 
\begin{equation}
\begin{split}
C_{\alpha\dot{\alpha}} (t,\vec{k})& =\int d^3xe^{-i\vec{k}\cdot\vec{x}}\langle\rho_{\alpha}(\vec{x},t)\bar{\rho}_{\dot{\alpha}}(\vec{0},0)\rangle=\\
& =-\bar\varepsilon e^{-D_sk^2t} [(i\gamma^1)_{\alpha\dot{\alpha}}\cos(kc_st)+(\hat{k}^j\gamma_j)_{\alpha\dot{\alpha}}\sin(kc_st)]
\end{split}
\end{equation}
We define the following correlator:
\begin{equation}
\tilde{C}_{\alpha\dot\alpha}(t,\vec x)=-i\theta(t)\langle\rho_{\alpha}(t,\vec x)\bar \rho_{\dot\alpha}(0,\vec 0)\rangle
\end{equation}
We can now calculate the Fourier transform:
\begin{equation}
\label{FT}
\begin{split}
\tilde C_{\alpha\dot{\alpha}}(\omega,\vec{k}) & =-i\int dt\int d^3xe^{i\omega t-i\vec{k}\cdot\vec{x}}\theta(t)\langle\rho_{\alpha}(t,\vec x)\bar \rho_{\dot\alpha}(0,\vec 0)\rangle\\
&=-i\int dt e^{i\omega t}\theta(t)C_{\alpha\dot{\alpha}} (t,\vec{k})\\
&=\bar\varepsilon\left[\frac{iD_sk^2+\omega}{(c_sk)^2+(D_sk^2-i\omega)^2}(i\gamma^1)_{\alpha\dot{\alpha}}+\frac{ic_sk}{(c_sk)^2+(D_sk^2-i\omega)^2}(\hat{k}^j\gamma_j)_{\alpha\dot{\alpha}}\right]
\end{split}
\end{equation}
The retarded correlator of the supercurrents is defined as:
\begin{equation}
C^R_{\alpha\dot\alpha}(t,\vec x)=-i\theta(t)\langle\{\rho_{\alpha}(t,\vec x),\bar \rho_{\dot\alpha}(0,\vec 0)\}\rangle
\end{equation}
From the spectral decomposition of $\tilde C$ and $C^R$ it follows:
\[\operatorname{Im}\tilde C_{\alpha\dot\alpha}(\omega,\vec k)=-\frac{\pi}{Z}\sum_{n,m}e^{-\beta E_n}\langle n| \rho_{\alpha}(0,\vec k)|m\rangle\langle m|\bar\rho_{\dot\alpha}(0,\vec 0)|n\rangle\delta(\omega+E_n-E_m)\]
\[\operatorname{Im}C^R_{\alpha\dot\alpha}(\omega,\vec k)=-\frac{\pi}{Z}\sum_{n,m}e^{-\beta E_n}(1+e^{-\beta \omega})\langle n| \rho_{\alpha}(0,\vec k)|m\rangle\langle m|\bar\rho_{\dot\alpha}(0,\vec 0)|n\rangle\delta(\omega+E_n-E_m)\]
Therefore we get:
\begin{equation}\label{relation}
\operatorname{Im}C^R_{\alpha\dot\alpha}(\omega,\vec k)=(1+e^{-\beta \omega})\operatorname{Im}\tilde C_{\alpha\dot\alpha}(\omega,\vec k)
\end{equation}
Taking the small $\vec k$ and then the small $\omega$ limit of (\ref{FT}) and using equation (\ref{relation}) we get: 
\begin{equation}
\bar\varepsilon D_s=\frac{1}{4}\lim\limits_{\omega \to 0}\left[\lim\limits_{k \to 0}\frac{\omega^2}{k^2}(i\gamma^1)^{\alpha\dot{\alpha}}\operatorname{Im}  C^{R}_{\alpha\dot{\alpha}}(\omega,\vec{k})\right]
\end{equation}
Using the continuity equation we can deduce a relation between the supercharge and the supercurrent correlators, namely:
\begin{equation}
\omega^2C^{R}_{\alpha\dot{\alpha}}(\omega,\vec{k})=k^2\hat{k}_i\hat{k}_jG^{ij}_{\alpha\dot{\alpha}}(\omega,\vec{k})=\frac{1}{3}k^2G^{ii}_{\alpha\dot{\alpha}}(\omega,\vec{k})
\end{equation}
where $ G^{ij}_{\alpha\dot{\alpha}}(t,\vec{x})=-i\theta(t)\langle\{S^i_{\alpha}(t,\vec{x}),\bar{S}^j_{\dot{\alpha}}(0,\vec 0)\}\rangle$ and in the last equality the $i$ indices are summed.
Using this relation we can write the Kubo formula in terms of the supercurrent retarded correlator:
 \begin{equation}
\label{Kubo}
\bar\varepsilon D_s=\frac{1}{12}\lim\limits_{\omega \to 0}\left[\lim\limits_{k \to 0}(i\gamma^1)^{\alpha\dot{\alpha}}\operatorname{Im}  G^{ii}_{\alpha\dot{\alpha}}(\omega,\vec{k})\right]
\end{equation}

\section{Normalization of the boundary action}
In this appendix we fix the normalization of the boundary action using the zero temperature calculation of the super stress energy tensor correlator.
Exceptionally in this appendix we use the conventions of \cite{Gates:1983nr}.\\
The leading term of the super energy-momentum tensor OPE has the following structure up to a constant $\mathcal{A}$ \cite{Anselmi:1996mq}:
\begin{equation}
J_{\alpha\dot{\alpha}}(z)J_{\beta\dot{\beta}}(0)=\mathcal{A}\frac{s_{\alpha\dot{\beta}}\bar{s}_{\beta\dot{\alpha}}}{(s^2\bar{s}^2)^2}
\end{equation}
where 
\begin{equation*}
\begin{split}
& s_{\alpha\dot{\alpha}}=(x-x')_{\alpha\dot{\alpha}}+\frac{i}{2}[\theta_{\alpha}(\bar{\theta}-\bar{\theta}')_{\dot{\alpha}}+\bar{\theta}'_{\alpha}(\theta-\theta')_\alpha]\\
&  \bar{s}_{\alpha\dot{\alpha}}=(x-x')_{\alpha\dot{\alpha}}+\frac{i}{2}[\bar{\theta}_{\dot{\alpha}}(\theta-\theta')_{\alpha}+\theta'_{\alpha}(\bar{\theta}-\bar{\theta}')_{\alpha}]\\
\end{split}
\end{equation*}
The super energy-momentum tensor is a superfield containing the axial current (at zeroth order in $\theta$) , the supercurrent (at linear order in $\theta$) and the energy-momentum tensor (at second order in $\theta$).
To fix the proportionality constant we can calculate the energy-momentum tensor correlator by applying twice the operator $\frac{1}{8}([\bar{D}_{\dot{\alpha}},D_{\alpha}]J_{\beta\dot{\beta}}+[\bar{D}_{\dot{\beta}},D_{\beta}]J_{\alpha\dot{\alpha}})$ on the super energy-momentum tensor OPE and compare with the energy momentum correlator \cite{Arutyunov:1999nw}:
\begin{equation}
T_{\mu\nu}T_{\rho\sigma}=\frac{5N_c^2}{\pi^4}\frac{J^{\kappa}_{\rho}J^{\lambda}_{\sigma}}{|x|^4}\left(\delta_{\mu\kappa}\delta_{\nu\lambda}+\delta_{\mu\lambda}\delta_{\nu\kappa}-\frac{1}{2}\delta_{\mu\nu}\delta_{\kappa\lambda}\right)
\end{equation}
where $J^{\nu}_{\mu}=\delta^{\nu}_{\mu}-2x_{\mu}x^{\nu}/ |x|^2$. From the comparison we deduce that:
\begin{equation}
\mathcal{A}=-\frac{32}{\pi^4}N^2_c
\end{equation}
Knowing the exact expression fot the super energy-momentum tensor OPE we can compute the supercurrent correlator by applying the operator $D_{\gamma}\bar D'_{\dot{\gamma}}$ and doing the proper symmetrizations.  Then we can compare with the zero temperature correlator of the supercurrent calculated in  \cite{Corley:1998qg}: 
\begin{equation}
S^{\alpha}_{\mu}\bar{S}^{\dot{\alpha}}_{\nu}=\frac{16\mathcal{N}}{\pi^2}\Pi_{\mu}^{\rho}\frac{\sigma^{\alpha\dot{\alpha}}_{\lambda}x^{\lambda}}{|x|^8}\left(\eta_{\rho\nu}-2\frac{x_{\rho}x_{\nu}}{|x|^2}\right)
\end{equation}
where $\Pi_{\mu}^{\rho}=\left(\delta_{\mu}^{\rho}-\frac{1}{4}\gamma_{\mu}\gamma^{\rho}\right)$ and find the value of the normalization of the action:
\begin{equation}
 \mathcal{N}=\frac{N_c^2}{\pi^2}
\end{equation}.

\end{document}